\documentclass{emulateapj}

\shorttitle{The Early Multi--Color Afterglow of GRB~050502a}
\shortauthors{Guidorzi et al.}

\begin{document}


\title{The Early ($<$1~hr) Multi--Color Afterglow of GRB~050502a:\\
Possible Evidence for a Uniform Medium with Density Clumps}

\author{C. Guidorzi, A. Monfardini\altaffilmark{1}, A. Gomboc\altaffilmark{2}, C.~G.~Mundell,
I.~A.~Steele, D.~Carter,\\ M.~F.~Bode, R.~J.~Smith,
C.~J.~Mottram, M.~J.~Burgdorf}
\affil{Astrophysics Research Institute, Liverpool John Moores University,
Twelve Quays House, Birkenhead, CH41 1LD, UK}

\email{(crg, am, ag, cgm, ias, dxc, mfb, rjs, cjm, mjb) @astro.livjm.ac.uk}


\author{N. R. Tanvir}
\affil{Centre for Astrophysics Research, University of Hertfordshire,
Hatfield AL10 9AB, UK}
\email{nrt@star.herts.ac.uk}

\author{N. Masetti}
\affil{INAF - Istituto di Astrofisica Spaziale e Fisica Cosmica, Sezione di
Bologna, Via Gobetti 101, I-40129 Bologna, Italy (formerly IASF/CNR, Bologna)}
\email{masetti@bo.iasf.cnr.it}

\and

\author{E. Pian}
\affil{Osservatorio Astronomico di Trieste, Via G.B. Tiepolo 11, 34131 Trieste, Italy.}
\email{pian@ts.astro.it}


\altaffiltext{1}{present address: ITC--IRST and INFN, Trento, via Sommarive, 18 38050 Povo (TN), Italy.}
\altaffiltext{2}{present address: FMF, University in Ljubljana, Jadranska 19, 1000 Ljubljana, Slovenia.}

\begin{abstract}
The 2-m robotic Liverpool Telescope reacted promptly to the gamma--ray burst
GRB~050502a discovered by {\em INTEGRAL} and started observing 3~min after the
onset of the GRB. The automatic identification of a bright afterglow of $r'\sim15.8$ triggered
for the first time an observation sequence in the $BVr'i'$ filters
during the first hour after a GRB. Observations continued for $\sim$1~day using
the {\em RoboNet-1.0} network of 2-m robotic telescopes.
The light curve in all filters can be described by a simple power law
with index of $1.2\pm0.1$. We find evidence for a bump rising at $t\sim0.02$~days
in all filters. From the spectrum and the light curve we investigate
different interpretative scenarios and we find possible evidence
for a uniform circumburst medium with clumps in density, as in the case of
GRB~021004. Other interpretations of such bumps, such as
the effect of energy injection through refreshed shocks or the result of a variable
energy profile, are less favored.
The optical afterglow of GRB~050502a is likely to be the result
of slow electron cooling with the optical bands lying between the synchrotron
peak frequency and the cooling frequency.
\end{abstract}





\section{Introduction}
\begin{figure}
\epsscale{.80}
\plotone{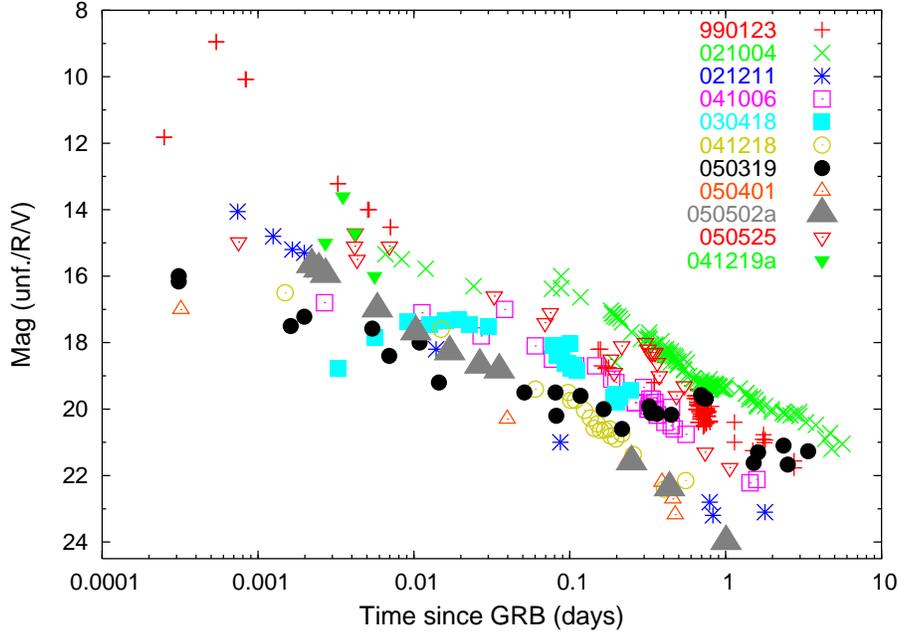}
\caption{Early light curves (unfiltered, $R$ and $V$) for a set of GRBs with detections
within minutes of the GRB. Grey triangles show the case of 050502a (filter
$r'$) robotically detected and followed-up by the Liverpool Telescope.
Data are taken from GCN circulars,
except for GRB~030418 \citep{Rykoff04} and GRB~041219a~\citep{Vestrand05}. Only the latter
values are corrected for Galactic dust extinction, which was high in this case ($\Delta R=4.9$).
\label{fig:all}}
\end{figure}

Although a considerable number of Gamma--Ray Bursts (GRBs) have
detected optical counterparts,
there are still few with optical afterglow measurements
within minutes of the gamma rays: Figure~\ref{fig:all} shows the early
light curves (unfiltered, $R$ and $V$) for all of these.
The early afterglow is particularly interesting as it carries information
about the immediate surroundings of the GRB progenitor, concerning either
the circumburst medium or the interaction between shells and the ISM in
the fireball scenario.
For two GRBs, an optical flash was detected
simultaneously with the gamma rays: GRB~990123 and
GRB~041219a: the former has been interpreted as the
signature of a reverse shock~\citep{Akerlof99}, while for the latter a correlation between
the gamma--ray and optical radiation light curves seems to favor a common origin
\citep{Vestrand05}.
These early afterglows show considerable variety: e.g., in the case
of GRB~030418 the optical emission was found to rise for the first 600~s,
slowly vary for 1400~s and then faded as a power law.
This was interpreted as due to the variable extinction by the local
circumburst medium \citep{Rykoff04}.
In the cases of GRB~990123 and GRB~021211, the early light curve is
described by a power law whose index varies from $\sim 2$ to
$\sim 1$ a few min after the GRB: at 0.5~min and 2.7~min in the
rest frame, respectively \citep{Holland04}. This has been interpreted as due
to the transition between reverse and forward shocks.

\placefigure{fig:all}

GRB~021004, one of the best observed GRBs in optical
\citep{Holland03,Fynbo05,deUgarte05}, exhibited a
number of bumps in its light curve, with all but the first bump
being detected from radio to U band.
Different interpretations have been suggested to explain the
light curve features: \citet{Lazzati02} modeled it using a variable density
profile, most likely a uniform medium with clumps with density variations
of the order of $\Delta n/n\sim10$ and size of $10^6$~cm.
Other authors \citep{Nakar03,Bjornsson04,deUgarte05} account for
the bumps with episodes of energy injections when inner shells catch up with
the afterglow shock at late times.
In addition, \citet{Nakar03} show that the bumps could be also explained by a
variable energy profile that is angularly-dependent on jet structure
(``patchy shell'' model).

In this Letter, we report the robotic detection and automatic
identification of GRB~050502a using the 2-m Liverpool Telescope
(LT) located in La Palma, Canary Islands: these observations
represent one of the first observations of a multi--color light curve in the first
hour since the burst. In addition, we report on late follow--up observations
performed with LT and the 2-m Faulkes Telescope North (FTN) located at Maui,
Hawaii, both members of the {\em RoboNet-1.0}
consortium\footnote{Funded by UK PPARC through a consortium of 10 UK universities.}
\citep{Gomboc05a}.

\section{Observations and Results}
On 2005 May 02 {\it INTEGRAL} detected GRB~050502a at 02:13:57 UT
and determined its position at $\alpha$=13:29:45.4
and $\delta$=+42:40:26.8 (J2000) with an error radius of 2~arcmin
(90\% C.L.) \citep{Gotz05_a}.
The GRB had a duration of 20~s. In the 20--200~keV band it had a peak
flux of $2\times10^{-7}$~erg~cm$^{-2}$~s$^{-1}$ and a fluence
of $1.4\times10^{-6}$~erg~cm$^{-2}$ \citep{Gotz05_b}, thus
ranking among faint/intermediate fluence GRBs.
ROTSE--IIIb started observing at 23.3~s after the GRB and
detected a 14.3-mag (unfiltered) unknown fading source at $\alpha$=13:29:46.3 and
$\delta$=+42:40:27.7 (J2000) ($l=98^{\circ}.76$, $b=+72^{\circ}.61$) \citep{Yost05}.
\citet{Prochaska05} acquired a spectrum with Keck--I 3.5~hr after
the GRB and identified a strong absorption feature, which they
interpret as SiII1260 at redshift $z=3.793$.

The LT responded robotically to the {\it INTEGRAL}
alert and started observing 3~min after the GRB onset (2.5~min after
the notice time).
Independently of ROTSE--IIIb it detected a bright fading source not
present in the USNO--B1.0, 2MASS and GSC~2.3 catalogs, with a
position consistent with that of the optical transient (OT) of ROTSE--IIIb
\citep{Gomboc05b}.
The automatic identification of the bright
and rapidly-fading OT by the LT GRB robotic pipeline (see \citet{Gomboc05c}
for technical details) resulted in the automatic triggering of a
multi--color imaging sequence that provided light curves in $BVr'i'$ filters
from 3~min to 1~hr after the GRB onset.
The robotic follow--up with LT ended after the first hour.
Subsequent follow--up observations were triggered manually on both the LT and FTN
(Table~\ref{tab:obs}).
Magnitudes in $r'$ and $i'$ have been calibrated using the SDSS~DR3
photometric database\footnote{http://cas.sdss.org/astro/en/tools/chart/navi.asp}.
We obtained a consistent calibration using Landolt standard field stars
\citep{Landolt92}, for which \citet{Smith02} provide SDSS calibration.
For the $B$ and $V$ filters, we calibrated with Landolt standard field
stars. The zero-points were stable during the night and fully consistent
with the photometric values. This is also confirmed by the Carlsberg Meridian
Telescope at La Palma\footnote{http://www.ast.cam.ac.uk/$\sim$dwe/SRF/camc\_extinction.html}.
Finally we corrected for the airmass and Galactic extinction.
The Galactic extinction \citep{Schlegel98} towards GRB~050502a is low:
$A_V=0.03$. We evaluated the extinction
in the other filters following \citet{Cardelli89}:
$A_B=0.04$, $A_{r'}=0.03$ and $A_{i'}=0.02$.
Magnitudes have been converted into flux densities $F_{\nu}$ (mJy)
following \citet{Fukugita95}.

\placetable{tab:obs}

Figure~\ref{fig:LC} shows the multi--color light curve acquired by the LT
during the first hour and the later points with both LT and FTN.
An achromatic bump rising at $t\sim0.02$~d is evident.
Fitting each light curve with a power law of the form $F\propto t^{-\alpha}$,
and excluding points $0.02$~d$<t<0.2$~d, we obtain power--law indices
consistent across all bands: $\alpha_B=1.20\pm0.04$,
$\alpha_V=1.16\pm0.06$, $\alpha_{r'}=1.19\pm0.04$, $\alpha_{i'}=1.16\pm0.03$.
By fitting only the $r'$ points obtained during
the detection mode within 3.8~min of the GRB onset time, we get
a power--law index of $\alpha_{r',\textrm{early}}=1.3\pm0.1$, consistent with 
the slopes reported above.

\placefigure{fig:LC}

Figure~\ref{fig:SED} shows the rest--frame Spectral Energy Distribution (SED) at
two epochs: before the bump ($t=0.004$~d), where 
no strong evidence for significant color change is observed (see Fig.~\ref{fig:LC}),
and at the bump ($t=0.035$~d).
Optical fluxes have been obtained by interpolation.
During the bump, a linear interpolation between consecutive points has
been adopted, considering that the variability timescales are much larger
than the time difference between the pairs of data points used for
interpolation.
Moreover, we back-extrapolated to $t=0.004$~d a {\it Swift} X--ray upper limit determined
around 1.3~d \citep{Hurkett05}, 
assuming a power--law decay, $F_X\propto t^{-\alpha_X}$,
and two different slopes: i) $\alpha_X=\alpha_X^{(1)}=1.45$ (solid arrow in Fig.~\ref{fig:SED});
ii) $\alpha_X=\alpha_X^{(2)}=0.95$ (dashed arrow in Fig.~\ref{fig:SED}).
The reasons for these choices are clarified in Sec.~\ref{sec:disc}.
In case (i) the power--law index between optical and X--rays must be: $\beta_{OX}>0.7$;
in case (ii) it must be: $\beta_{OX}>1.1$.
However a word of caution is needed,
particularly because we know from the {\it Swift} observation that during the first
few hundred seconds the early X--ray afterglows can be characterized by
a steep decline followed by a shallower decay \citep{Tagliaferri05}.
The back-extrapolation for the radio upper limits provided
by \citet{vanderHorst05} between 0.6~d and 1.1~d is much more difficult, given that
in general the behavior of the early radio afterglow is likely to be very different
from the optical one. Hereafter, we do not consider these radio limits.

\placefigure{fig:SED}

We note a possible marginal reddening of the spectrum
at the time of the bump (see bottom panel of the inset in Fig.~\ref{fig:SED}),
albeit not statistically significant: the flux ratio between
the bump and the pre-bump epochs does not vary significantly for different optical
bands (see also GRB~000301C, Masetti et al. 2000).\nocite{Masetti00}
Due to the high $z$, the Lyman-$\alpha$ forest suppresses both
$B$ and $V$ band fluxes. This accounts for the unusually-steep SED in the optical:
by fitting all the four points with a power law, $F\propto \nu^{-\beta}$, the
index is around $\beta=2.8\pm0.8$ with a poor $\chi^2$ ($\chi^2/{\textrm dof}=116/2$).
However, if we assume a standard value
of $\beta=0.8$ (see Sec.~\ref{sec:disc}),
we find that the flux deficiency at high $\nu$ can be ascribed to
the Lyman-$\alpha$ forest (see the top panel of the Inset in Fig.~\ref{fig:SED}). 

\section{Discussion}
\label{sec:disc}
\begin{figure}
\epsscale{.80}
\plotone{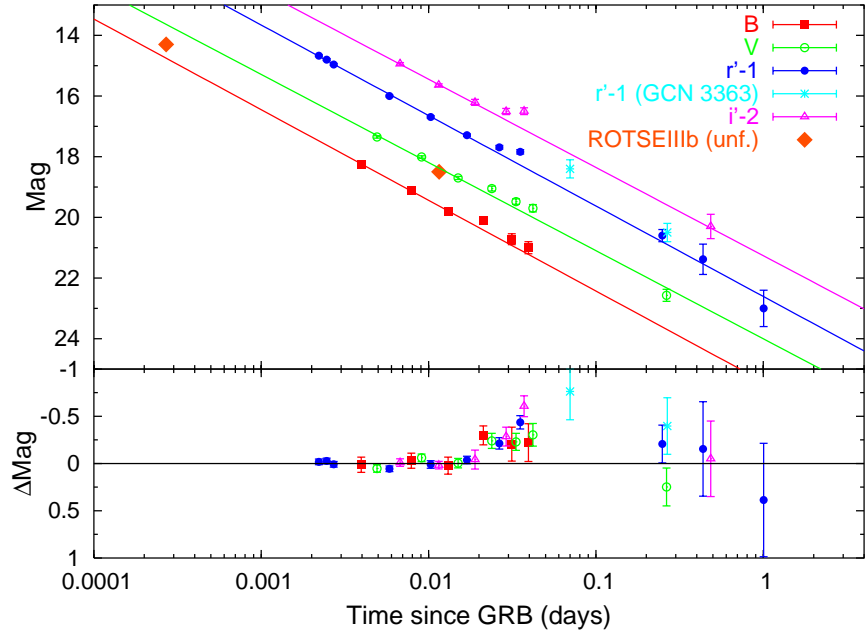}
\caption{{\em Top Panel}: Multi--color light curve of GRB~050502a measured with the Liverpool
and the Faulkes North Telescopes. Also
shown are the best-fit power laws: all of them are consistent with a power--law
index of $1.2\pm0.1$ (see text). Two ROTSE--IIIb unfiltered points \citep{Yost05} and two $r'$
points derived from \citet{Mirabal05} are plotted as well.
{\em Bottom Panel}: residuals with respect to the best-fitting power laws.\label{fig:LC}}
\end{figure}

The reality of the bump we find in the light curve at $t\sim0.02$~d 
is also supported by a rebrightening observed in the IR \citep{Blake05}: initially
they observed a decay of 1.1~mag in the $J$ band between 47~min and 94~min (corresponding
to a power--law decay index of $\alpha=1.5$, no error reported), followed by
a rebrightening of $\Delta J\sim0.1$ between 94~min (0.065~d) and 121~min (0.084~d).
In addition to our measurements, Fig.~\ref{fig:LC} also shows two unfiltered points by
ROTSE--IIIb \citep{Yost05} and two other $R$ measures reported by \citet{Mirabal05},
which we converted to $r'$ assuming $0.3<R-I<0.6$ (no uncertainty was reported,
so we assumed the systematic of 0.3 of the USNO--B1.0 magnitudes, as they calibrated
with a USNO--B1.0 field star). In particular, the latter points seem to confirm the
presence of the bump in $r'$, despite the large uncertainties.
\citet{Durig05} report unfiltered observations of the bump.  Since the
conversion of unfiltered to standard magnitudes requires some assumptions
and implies large uncertainties, we are not as confident about the proper
intercalibration of those converted magnitudes and our data as we are at
earlier epochs, when the decay is simply monotonic. Therefore, lacking a
comparison dataset of unfiltered data covering both the monotonic early
decay and the bump, we have not included \citet{Durig05} data in Fig.~\ref{fig:LC}.

Following \citet{Lazzati02}, if we interpret the bump as due to density variations
of the ISM, this is possible only if the observation occurred at a frequency $\nu=\nu_O$
(let $\nu_O$ be the frequency of our optical bands) below the cooling break $\nu_c$
and above the peak synchrotron frequency $\nu_m$: $\nu_m<\nu<\nu_c$.
In the following we consider the two cases of uniform ISM and wind environment, respectively.

\begin{figure}
\epsscale{.80}
\plotone{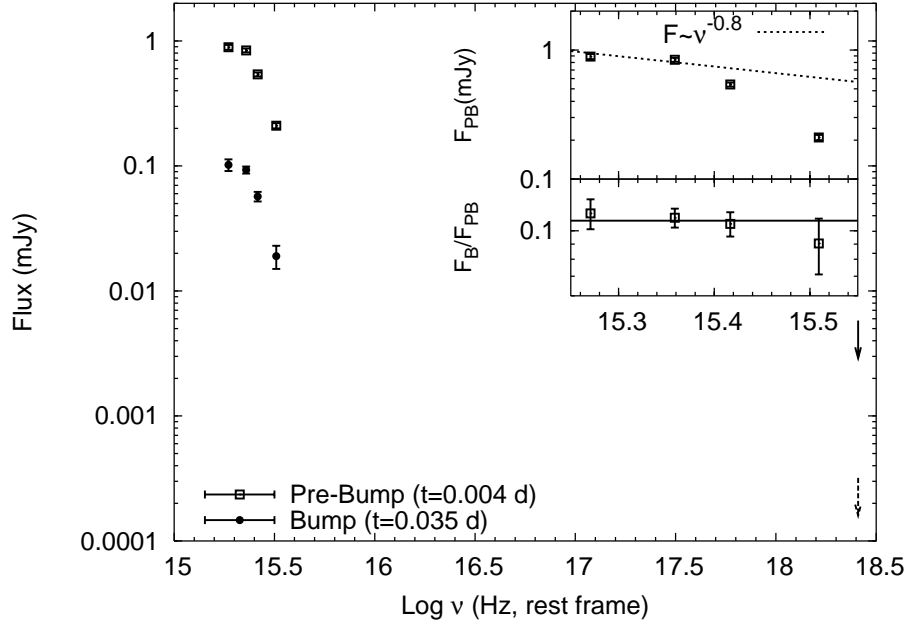}
\caption{Rest--frame SED at two epochs: $t=0.004$~d (Pre-Bump) and $t=0.035$~d (Bump).
Optical points have been interpolated at the same epochs. The X--ray upper limit at $t=0.004$~d
(solid arrow) has been obtained by back-extrapolating the values provided by \citet{Hurkett05},
around $\sim$1.3~d, assuming a power--law decay with index of $\alpha_X=1.45$. Alternatively,
the other X--ray upper limit at $t=0.004$~d (dashed arrow) is obtained assuming $\alpha_X=0.95$ (see text).
{\it Inset, top panel}: close-up of the Pre-Bump optical points 
with the power law with $\beta=0.8$ (dotted line). The flux deficiency at high $\nu$ is
due to the Lyman-$\alpha$ forest (see text).
{\it Inset, bottom panel}: flux ratio between the Bump and the Pre-Bump epochs as a function
of $\nu$. All the ratios are consistent with a constant value (weighted average of $0.108\pm0.005$,
$\chi^2/{\textrm dof}=1.2$) shown by the solid line.\label{fig:SED}}
\end{figure}

In the case of uniform ISM, the expected power--law index of the light
curve is $\alpha=3(p-1)/4$, where $p$ is the electron energy distribution index
\citep{Sari98}. From our measure of $\alpha=1.2\pm0.1$ we derive $p=2.6\pm0.1$.
We also note that when $\nu_c$ crosses the optical band we should expect a steepening
in the light curve of $\Delta\alpha=0.25$. Since we do not find evidence for
this before $t<1$~d, the only possibility is that $\nu_O<\nu_c$ at least until $t\sim$1~d.
The energy spectrum at frequency $\nu_m<\nu<\nu_c$ is a power law with index
$\beta=(p-1)/2$, i.e. $\beta=0.8\pm0.05$.
Figure~\ref{fig:SED} shows that this is consistent with our result.
The cooling break $\nu_c$ must lie between the optical band $\nu_O$ and the X--ray
$\nu_X$: $\nu_O<\nu_c<\nu_X$. The power--law index of the spectrum between $\nu_c$ and
$\nu_X$ is expected to be $\beta_{cX}=p/2=1.3\pm0.05$.
The X--ray power--law decay index, $\alpha_X$, is expected to be:
$\alpha_X=3(p-1)/4$ ($\nu_c>\nu_X$), $\alpha_X=(3p-2)/4$ after $\nu_c$ has crossed
the X--ray band ($\nu_c<\nu_X$), thus experiencing a steepening of $\Delta\alpha_X=0.25$.
As this is expected to occur soon after the GRB, it is sensible to back-extrapolate
the X--ray upper limit assuming for most of the time $\alpha_X=(3p-2)/4=1.45$.
From Fig.~\ref{fig:SED}, as long as we assume the validity of the
X--ray upper limit back-extrapolated to $t=0.004$~d assuming $\alpha_X=1.45$ (solid
arrow), we find that the shallowest power--law index allowed between optical and X--rays is $\beta_{OX}>0.7$ .
Thus, this is consistent with a broken power law with power--law indices from $0.8$ to $1.3$.
In summary, we conclude that the case of a uniform ISM is fully consistent
with our observations.

In the case of wind environment and $p<2$ we must use the relation $\alpha=(p+8)/8$ by
\citet{Dai01}  for $\nu_m<\nu<\nu_c$, which yields $p=1.6\pm0.8$. The case of $p>2$ is
incompatible with the data: from the relation $\alpha=(3p-1)/4$ by \citet{Chevalier99} we derive
a value of $p=1.9\pm0.1$.
From $\beta_{mc}=(p-1)/2$ and $\beta_{cX}=p/2$,
holding for $\nu_m<\nu<\nu_c$ and for $\nu_c<\nu<\nu_X$, respectively, 
we derive: $\beta_{mc}=0.3\pm0.4$ and $\beta_{cX}=0.8\pm0.4$.
Concerning the back-extrapolation of the X--ray upper limit, $\alpha_X$ is expected
to be: $\alpha_X=(p+8)/8$ ($\nu_c>\nu_X$), $\alpha_X=(p+6)/8$ after $\nu_c$ has crossed
the X--ray band ($\nu_c<\nu_X$), thus experiencing a steepening of $\Delta\alpha_X=0.25$.
For the same reason as in the previous case, it is reasonable to assume $\alpha_X=(p+6)/8=0.95$
for most of the time. The consequent limit on the spectrum is $\beta_{OX}>1.1$ (dashed arrow
in Fig.~\ref{fig:SED}).
This is compatible only with $\beta_{cX}$.
Furthermore, $\nu_c$ should be very close to the
optical bands: this implies that during our observation $\nu_c$ should cross the optical
bands, producing a slope change in the power--law decay of $\Delta\alpha=0.25$, which is
not observed.
If we assume that $\nu_c>\nu_X$ for most of the time between
$t=0.004$~d and the epoch of the X--ray observation ($\sim1.33$~d), we derive the X--ray
upper limit assuming $\alpha_X=(p+8)/8=1.2$, yielding $\beta_{OX}>0.9$, which is not
consistent with $\beta_{OX}=\beta_{mc}=0.3\pm0.4$.

In contrast to GRBs 990123 and 021211, we find no evidence for a change in the temporal
slope within the first few minutes of the onset of GRB~050502a, ruling out
a transition from reverse to forward shock emission at this time. In GRB~050502a
the bump rises at $\sim$6~min after the GRB in the rest-frame, to be compared with
0.5~min and 2.7~min of GRB~990123 and GRB~021211, respectively, when the above
transition between reverse and forward shocks is supposed to occur.
Should GRB~050502a have exhibited a similar transition, we should have detected
it before the bump.
We conclude that, despite the fact that a wind environment cannot be ruled out,
the uniform ISM with clumps in density seems to better account for our observations.

The interpretation of the bump as the result of a refreshed shock
catching up with the afterglow front shock seems more problematic, even if it
cannot be ruled out. In fact, according to the original refreshed-shocks scenario
\citep{Kumar00,Granot03}, we should expect that the duration $\Delta t$ of the bump is
comparable with its start time: $\Delta t\approx t$. In the case
of GRB~050502a our measures and those by \citet{Mirabal05} show that,
in spite of the uncertainty, $\Delta t\approx0.2$~d and $t\sim0.02$~d.
Following \citet{Kumar00}, the impact between the two shells should
produce a forward shock in the outer shell responsible for the
bump and a reverse shock propagating in the inner shell. If $E_1$ and $E_2$
are the energy of the outer and inner shells, respectively, the increase in the
emission due to the forward shock is expected to be $f=(1+E_2/E_1)^{(p+3)/4}$.
From Fig.~\ref{fig:LC} we measure a flux increase of $10^{\Delta m/2.5}\sim1.6$
($\Delta m\sim0.5$); from $p=2.6$ we obtain $E_2/E_1\sim0.4$.
The spectrum at the bump is expected to have two peaks: the lower $\nu$ peak
is due to the reverse shock in the inner shell and its frequency should be
$\sim7\gamma_{0i}^2(E_2/E_1)^{1.1} \simeq 64(\gamma_{0i}/5)^2$ times lower than
the peak frequency of the outer shell, i.e. $\nu_m$, which we know is below
the optical bands at the time of the bump ($\gamma_{0i}$ is the Lorentz factor
of the outer shell at the time of impact). The increase of emission at this
frequency due to the inner shell is expected to be a factor
$\sim8(\gamma_{0i}E_2/E_1)^{5/3} \simeq 25(\gamma_{0i}/5)^{5/3}$.
Thus, the bump should have been more evident at low frequency: $\nu_m/64/(\gamma_{0i}/5)^2<\nu_O$,
i.e. IR or radio. Unfortunately, the lack of early radio
observations prevents this prediction from being tested.
In the $J$-band \citet{Blake05} report a
rebrightening of $\sim0.1$~mag, which however seems smaller than that observed
by us in the optical. Moreover, according to \citet{Blake05} the $J$-band
rebrightening occurs between 0.065~d and 0.084~d, i.e. later than 0.02~d of
the optical bands.

In conclusion, although the refreshed-shock scenario cannot be completely
ruled out due to the lack of early radio observations, our observations
appear to be more difficult to reconcile with its predictions than with
those of the variable density environment.

\acknowledgments
CG and AG acknowledge their Marie Curie Fellowships from the European Commission.
CGM acknowledges financial support from the Royal Society.
AM acknowledges financial support from PPARC.
MFB is supported by a PPARC Senior Fellowship.
The Liverpool Telescope is operated on the island of La Palma by
Liverpool John Moores University at the Observatorio del Roque de los
Muchachos of the Instituto de Astrofisica de Canarias.
The Faulkes Telescope North is operated with support from the Dill
Faulkes Educational Trust.

\clearpage

\clearpage

\begin{deluxetable}{llrrrl}
\tabletypesize{\scriptsize}
\tablecaption{Optical Photometry for GRB~050502a with LT and FTN\label{tab:obs}}
\tablewidth{0pt}
\tablehead{
\colhead{Telescope} & \colhead{Filter} & \colhead{Start}\tablenotemark{a} & \colhead{Exposure} & \colhead{Mag.} & \colhead{Comment}\\
                &                 & \colhead{(min)} & \colhead{(s)}      &                     &\\
}
\startdata
LT & SDSS-R &     3.1 &  10 & $15.67\pm0.03$ & detection mode\\
LT & SDSS-R &     3.5 &  10 & $15.80\pm0.03$ & detection mode\\
LT & SDSS-R &     3.8 &  10 & $15.96\pm0.03$ & detection mode\\
LT & Bessell-B &  5.4 &  30 & $18.25\pm0.08$ & multi-color sequence\\
LT & Bessell-V &  6.7 &  30 & $17.35\pm0.04$ & multi-color sequence\\
LT & SDSS-R &     8.1 &  30 & $17.00\pm0.03$ & multi-color sequence\\
LT & SDSS-I &     9.5 &  30 & $16.94\pm0.04$ & multi-color sequence\\
LT & Bessell-B & 10.8 &  60 & $19.10\pm0.08$ & multi-color sequence\\
LT & Bessell-V & 12.6 &  60 & $18.01\pm0.04$ & multi-color sequence\\
LT & SDSS-R &    14.3 &  60 & $17.69\pm0.04$ & multi-color sequence\\
LT & SDSS-I &    16.1 &  60 & $17.64\pm0.04$ & multi-color sequence\\
LT & Bessell-B & 17.8 & 120 & $19.81\pm0.09$ & multi-color sequence\\
LT & Bessell-V & 20.6 & 120 & $18.70\pm0.05$ & multi-color sequence\\
LT & SDSS-R &    23.4 & 120 & $18.29\pm0.04$ & multi-color sequence\\
LT & SDSS-I &    26.2 & 120 & $18.21\pm0.10$ & multi-color sequence\\
LT & Bessell-B & 29.1 & 180 & $20.12\pm0.10$ & multi-color sequence\\
LT & Bessell-V & 32.9 & 180 & $19.05\pm0.08$ & multi-color sequence\\
LT & SDSS-R &    36.6 & 180 & $18.69\pm0.06$ & multi-color sequence\\
LT & SDSS-I &    40.4 & 180 & $18.51\pm0.10$ & multi-color sequence\\
LT & Bessell-B & 44.2 & 120 & $20.72\pm0.18$ & multi-color sequence\\
LT & Bessell-V & 47.0 & 120 & $19.48\pm0.09$ & multi-color sequence\\
LT & SDSS-R &    49.8 & 120 & $18.84\pm0.07$ & multi-color sequence\\
LT & SDSS-I &    52.6 & 120 & $18.50\pm0.11$ & multi-color sequence\\
LT & Bessell-B & 55.3 & 180 & $21.00\pm0.20$ & multi-color sequence\\
LT & Bessell-V & 59.1 & 180 & $19.70\pm0.12$ & multi-color sequence\\
FTN& Bessell-R &  348 & 4x200& $21.6\pm0.2$  & late follow-up\\
FTN& Bessell-V &  370 & 6x200& $22.6\pm0.2$  & late follow-up\\
FTN& Bessell-R &  620 & 4x200& $22.4\pm0.5$  & late follow-up\\
FTN& SDSS-I &     690 & 4x200& $22.3\pm0.4$  & late follow-up\\
LT & SDSS-R &    1340 & 24x150& $24.0\pm0.6$ & late follow-up\\
\enddata



\tablenotetext{a}{This corresponds to the time delay with
respect to the GRB trigger time, $t_0=0.09302$~UT.}.

\end{deluxetable}






\end{document}